\documentclass[12pt,draftcls,onecolumn]{IEEEtran}
\usepackage[T1]{fontenc}
\usepackage[latin9]{inputenc}
\usepackage{array}
\usepackage{verbatim}
\usepackage{float}
\usepackage{textcomp}
\usepackage{amsmath}
\usepackage{subfigure}
\usepackage{amsthm}
\usepackage{graphicx}
\usepackage{amssymb}
\usepackage{esint}
\usepackage{algorithmic}

\makeatletter

\newcommand{\lyxmathsym}[1]{\ifmmode\begingroup\def\b@ld{bold}
  \text{\ifx\math@version\b@ld\bfseries\fi#1}\endgroup\else#1\fi}

\providecommand{\tabularnewline}{\\}
\floatstyle{ruled}
\newfloat{algorithm}{tbp}{loa}
\floatname{algorithm}{Algorithm}

\newtheorem{thm}{Theorem}
\newtheorem{cor}{Corollary}
\newtheorem{prop}{Proposition}
\newenvironment{lyxlist}[1]
{\begin{list}{}
{\settowidth{\labelwidth}{#1}
 \setlength{\leftmargin}{\labelwidth}
 \addtolength{\leftmargin}{\labelsep}
 }}
{\end{list}}

\makeatother

\makeatother

\makeatother

\makeatother

\makeatother

\makeatother

\makeatother

\makeatother

\makeatother

\makeatother

\makeatother

\makeatother

\makeatother

\makeatother

\begin{document}

\title{Angle Tree: Nearest Neighbor Search in High Dimensions with Low Intrinsic
Dimensionality}

\author{Ilia Zvedeniouk and Sanjay Chawla  \\
School of Information Technologies \\
University of Sydney\\}
\maketitle
\begin{abstract}
We propose an extension of tree-based space-partitioning indexing
structures for data with low intrinsic dimensionality embedded in
a high dimensional space. We call this extension an Angle Tree. Our
extension can be applied to both classical kd-trees as well as the
more recent rp-trees.

The key idea of our approach is to store the angle (the {}``dihedral
angle'') between the data region (which is a low dimensional manifold)
and the random hyperplane that splits the region (the {}``splitter'').

We show that the dihedral angle can be used to obtain a tight lower
bound on the distance between the query point and any point on the
opposite side of the splitter. This in turn can be used to efficiently
prune the search space. We introduce a novel randomized strategy to
efficiently calculate the dihedral angle with a high degree of accuracy.

Experiments and analysis on real and synthetic data sets shows that
the Angle Tree is the most efficient known indexing structure for
nearest neighbor queries in terms of preprocessing and space usage
while achieving high accuracy and fast search time.

\end{abstract}

\section{Introduction }

\begin{table}
\begin{centering}
 \begin{tabular}{|>{\raggedright}p{0.67in}|>{\raggedright}p{0.7in}|>{\raggedright}p{0.75in}|>{\raggedright}p{0.7in}|>{\raggedright}p{0.63in}|}
\hline 
 & Preprocessing & Space & Search Time & Accuracy\tabularnewline
\hline
\hline 
Brute Force & None & None & V.Bad & Exact\tabularnewline
\hline 
Cover tree & Bad(3)%

{} & Good(2) & Good(=2) & Exact(1) \tabularnewline
\hline 
LSH & Good(2) & Bad(3)%

{} & V.Good(1) & High(=2) \tabularnewline
\hline 
\textbf{Angle Tree} & \textbf{V.Good(1)} $\thickapprox kd-tree$  & \textbf{V.Good(1)}
 $\thickapprox kd-tree$ & \textbf{Good(=2)} & \textbf{High(=2)} \tabularnewline
\hline
\end{tabular}

{\small ~}\\
{\scriptsize \caption{{\small Qualitative evaluation of the current state of the art NNS
data structures, brute force, and the Angle Tree. (\#) is the rank
out of the three data structures, 1 being the best. See Section
\ref{sec:exp} for more details.}}
}{\scriptsize \par}

\end{centering}
\end{table}

{}The Nearest Neighbor Search (NNS) problem is of fundamental importance
with wide applicability in Search, Pattern Recognition and Data Mining.
The problem is simply defined as: \\

\textit{Given a data set $S$ of size $N$ and dimension $D$, efficiently
preprocess $S$ so that given a query point $q$, we can quickly find
the nearest points to $q$ (nearest neighbors) in $S$.} \\

Without any preprocessing of $S$ the brute force time complexity
of NNS is $O(ND)$. This is impractical for very large databases.

In low dimensional space, data structures like kd-trees \cite{friedman}
are very efficient resulting in expected time complexity of $O(DlogN)$.
A long standing open problem is to design data structures which can
scale up to high dimensions because experience shows that in high
dimensions, space partitioning data structures become inefficient.
For example, kd-trees require that $N\gg2^{D}$ or else search performance
degenerates to brute force \cite{wikiIndyk}. This has been the status quo for some time.

Tenenbaum et al. \cite{tanen} introduced a novel approach to analysing
low dimensional manifolds embedded in high dimensions. This approach
is widely applicable as many real world high-dimensional data sets
only have a small number of non-linear degrees of freedom. For example,
consider a collection of images of human faces. The variance of this
data set can be described by considering the size of nose, distance
between eyes, and a small number of other such degrees of freedom.
Each image can be described by these degrees of freedom much more
succinctly than by considering the thousands of pixels that make up
the image.

Any data set containing images (or other kinds of sampling of the
real world, which is usually fairly ordered) cannot be completely
{}``random''. An example of a near random data set would be images
of a television snow screen.

\begin{figure}
\noindent \begin{centering}
\includegraphics[scale=0.5]{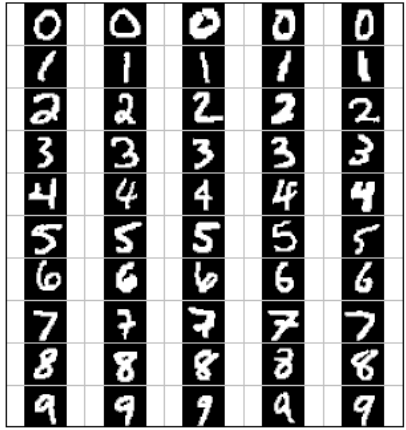}
\par\end{centering}

\caption{mnist dataset. The digits vary in a limited number of ways; inclination,
size, loop articulation etc. Hence despite this data set having a high ambient dimensionality 
(number of pixels in each image) the intrinsic dimensionality is low.}

\end{figure}

Many real-world data sets are likely to be reducible to a much lower
dimension, where each axis represents a non-linear degree of freedom.
We do not propose to find these intrinsic degrees of freedom. We only
assume their existence in the data set in question. Several definitions
of intrinsic dimensionality have been proposed, such as small covering
numbers, Assouad dimension, low dimensional manifolds, and Local Covariance Dimension (LCD).
In this paper we mostly consider LCD.

Random projection trees (rp-trees) were introduced by \cite{Dasgupta,Freund}.
This data structure differs from the kd-tree only in the nature of
the splitter used. While the kd-tree always splits parallel to a single
axis, the rp-tree splits in essentially a random direction. \cite{Dasgupta,Freund}
show that the rp-tree adapts to the intrinsic dimensionality of data
(in terms of grouping nearby points in the same cells) much better
than the kd-tree. However, NNS (using rp-trees) with the standard
kd-tree pruning method still degenerates to brute force search in
high dimensions.

The Cover Tree\cite{Beygel2006} is a new non-space partitioning NNS
data structure that exploits the intrinsic dimensionality of data.
It achieves very good search time and space usage, but suffers from
slow preprocessing \cite{Shi06,Comparison2005}. The preprocessing
requirements are also very sensitive to noise\cite{Beygel2006}. For
this reason, it is not practical for use on large, high-dimensional
and noisy data, such as image databases.

The most popular current high-dimensional NNS technique is Locality-Sensitive
Hashing. It is almost practical for the real world, except for its
large space usage\cite{multi-probe,Joly2008}. The index it builds
is often many times larger than the original data, again making it
unsuitable for very large databases.

The approximate NNS technique described in \cite{flann} proposes two
solutions. One is hierarchical k-means trees, which is
an example of a Geometric Nearest-Neighbor Access Tree (GNAT) \cite{gnat}. 
These trees can suffer from slow preprocessing.
The other solution is multiple randomized kd-trees. Since searching through 
kd-trees in high dimensions degenerates to  brute force, \cite{flann} obtain 
approximate seach by simply searching through a fixed number of leaf nodes
 for each tree. This is similar in spirit to multi-probe LSH, and as 
 we discuss in Section \ref{sec:disc}, this approach is very likely to be 
improved by combining it with our method.

The process described in \cite{tanen} is also not practical for large,
high-dimensional data sets as it requires an eigendcomposition of
the gram matrix - an $O(n^{3})$ operation. However, to perform efficient
NNS, we will show that the expensive process of manifold learning
can be avoided.

\subsection*{Contribution}

We propose a data structure (Angle Tree) that for the first time gives
us the ability to index large and high-dimensional data sets in practice.
Its preprocessing and space requirements are not much greater than
a regular kd-tree, while providing fast and highly-accurate NNS capability.
It also maintains the simplicity of the original kd-tree.

As we will show, in our implementation using a space partitioning
data structure, all we need to extract from the data are the angles
that the data region in question makes with the splitting hyperplane
(splitter). This angle has meaning even in high-dimensional space
by the definition of the dot product.

The Angle Tree is described in Algorithms 1, 2 and 3, and is included
in the Appendix. It is conceptually a small modification to any current
tree-based indexing technique, such as a kd or rp-tree\cite{Dasgupta,Freund},
that adds the power to exploit the intrinsic structure of data during
NNS, while introducing a very small probability of error (not finding
the true nearest neighbors). While these tree structures can already
perform near-neighbor search (ie. only search one cell of the tree),
we introduce a new method of efficiently deciding which other cells,
if any, also need to be searched. This method is a generalization
of the classic kd-tree NNS algorithm for low intrinsic dimensionality
data embedded in very high-dimensional space. If there is no intrinsic
structure to exploit in a particular data set, then the Angle Tree
will behave like a regular kd or rp-tree, with minimal additional
overhead.

The remainder of the paper is organized as follows. In Section \ref{sec:def} we
list all definitions used in the paper. In Section \ref{sec:keyidea} we 
present the key idea of this paper and prove its correctness. In Section 
\ref{sec:highDimBehav} we discuss the behaviour of the dihedral angle (between the splitter 
and the low dimensional manifold region) in high dimensions. In Section \ref{sec:estimDihed} we introduce our method for finding the dihedral angle. In section
 \ref{sec:errorReg} we analyze the accuracy of NNS using Angle Trees. In Section \ref{sec:exp} we report the results of our experiments. In Section \ref{sec:disc} 
we provide some further comments and suggest directions for future
work.

\section{Definitions} \label{sec:def}

In this section we collect all the definitions used in the paper for
the convenience of the reader.
\begin{itemize}
\item \textbf{Ambient Dimension ($D$):} is the dimension of the raw data.
For example, if data is presented as an $N\times D$ matrix, then
the ambient dimension is $D$. 
\item \textbf{Intrinsic Dimension ($d$):} is the number of degrees of freedom
in the data. Again, if data is presented as an $N\times D$ matrix
then the intrinsic dimensionality ($d$) is typically much smaller
than $D$. The intrinsic dimension is usually not known but can be
estimated \cite{pami1995}. 
\item \textbf{Local Covariance Dimension (LCD):} {}A set S has local covariance
dimension $(d,\epsilon,r$) if it has $(1\lyxmathsym{\textminus}\epsilon)$
fraction of its variance concentrated in a $d$-dimensional subspace.
More precisely, let $\sigma_{1}^{2},\sigma_{2}^{2},...,\sigma_{D}^{2}$
denote the eigenvalues of the covariance matrix; these are the variances
in each of the eigenvector directions. "Set $S\subset\mathbb{R}^{D}$
has local covariance dimension $(d,\epsilon,r$) if its restriction
to any ball of radius $\le r$ has covariance matrix whose largest
$d$ eigenvalues satisfy $\sigma_{i}^{2}\ge\frac{1-\epsilon}{d}\left(\sigma_{1}^{2},+\sigma+...+\sigma_{D}^{2}\right)$"
\cite{Dasgupta} 
\item \textbf{Intrinsic Plane (IP):} is the $d$-dimensional affine subspace
associated with the LCD defined above. 
\item \textbf{Splitter:} the splitting hyperplane of dimension $D-1$ to
$\mathbb{R}^{D}$. It splits the data region in question into two
not necessarily equal parts. 
\item \textbf{Dihedral Angle ($\alpha$):} The angle between the splitter
and the IP. If the IP has codimension 1, then this 
angle is simply the angle between their normal vectors (the splitter 
always has codimension of 1). Otherwise, it is $\pi/2$ minus the angle between 
the normal vector to the splitter and its projection onto the IP \cite{dihed}. 
\item \textbf{Error Angle ($\theta$):} This angle reflects the accuracy
of the dihedral angle returned by the getAngle() function (see Section \ref{sec:estimDihed}). 
\item \textbf{Ambient Distance ($dist(q,p)$):} is the distance in the ambient
space between points or hyperplanes $q$ and $p$. 
\item \textbf{Manifold Distance ($dist_{M}(q,p)$):} is the distance between
points or hyperplanes $q$ and $p$ when the path is restricted to
the data manifold. 
\item \textbf{Intrinsic Plane Distance ($dist_{IP}(q,p)$):} is the distance
between points or hyperplanes $q$ and $p$ when the path is restricted
to the IP. If $q$ or $p$ does not lie on the IP, then we project it
onto the IP before calculating the distance. Note that $dist_{m}(q,p)\ge dist(q,p)\ge dist_{IP}(q,p)$
if $p$ and $q$ are both points. 
\item \textbf{Locality-Sensitive Hashing (LSH): }A popular current NNS solution
proposed by \cite{Indyk04,Andoni2009}. In this paper we refer
to the random projection implementation of LSH. Its main problem is
space usage \cite{multi-probe}. 
\item \textbf{Cover Tree: }A relatively new NNS tree structure
that is designed to exploit $d$. It does not partition the space, and
in fact only requires a metric that satisfies the triangle inequality.
Its main problem is preprocessing time\cite{Shi06,Comparison2005}
and sensitivity to noise \cite{Beygel2006}. 
\item \textbf{Random Projection Tree (rp-tree):} Data structure introduced
by \cite{Dasgupta,Freund}. Similar to kd-tree, but splits in a random
direction. Has the property that every $O(dlogd)$ levels, the diameter
(distance between two furthest points) of each cell is halved \cite{Dasgupta,Freund}. 
\end{itemize}

\section{Key Idea} \label{sec:keyidea}

\begin{figure}
\begin{center}
\subfigure[kd-tree NNS Search]{\label{fig:search-a} \includegraphics{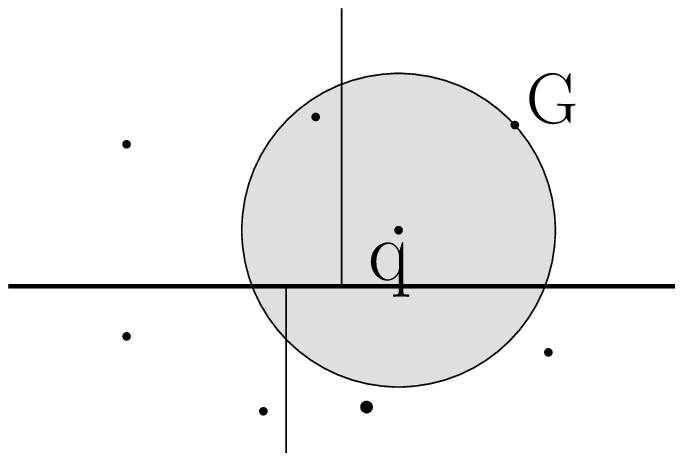}}
\subfigure[Manifold Search]{\label{fig:mainidea} \includegraphics{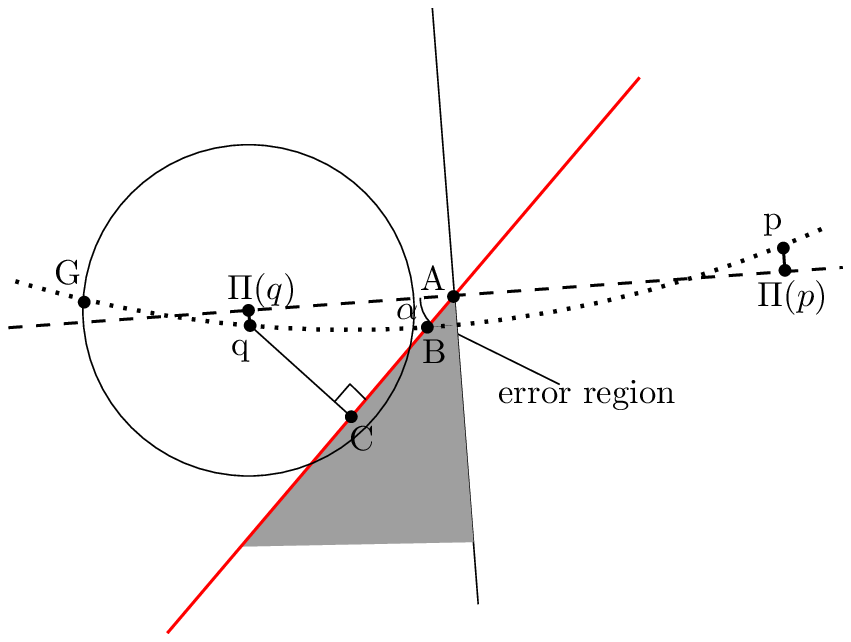}}
\end{center}
\caption{kd-tree NNS search ball. Q is query point. G is the nearest neighbor
found so far. The top level split is in bold.}
\label{fig:search}
\end{figure}

In order to appreciate the key insight of the proposed approach we first
explain how a query point $q$ uses the classical kd-tree to carry out the NNS.

A query point $q$ will navigate down a branch of the kd-tree (each
pair of branches being split by a splitter) until it reaches the
leaf cell in which it is contained. The nearest neighbor in that cell
will be identified ($G$ in Figure \ref{fig:search-a}). A search sphere
of radius $dist(q,G)$ will be constructed and any neighboring cell which
intersects the search sphere will be explored for points which are
potentially closer to $q$ than $G$.  Effectively this implies that
the perpendicular distance between $q$ and the splitter(s), which form the
walls of the cell, serve as a lower bound for pruning. 

It is well known that in high-dimensional space, the kd-tree NNS degenerates
to little better than brute force. We will provide an alternate explanation
 in Section \ref{sec:highDimBehav} in terms of the distance from $q$ to the walls of its cell.

Now if the data lives in a low dimensional manifold, for example, on
the dotted line in Figure \ref{fig:mainidea} then a better lower bound
is the manifold distance $dist_M(q,B)$. However, generally the manifold
is not known, so we instead approximate $dist_M(q,B)$ with the IP distance
$dist_IP(q,A)$. 

In order to calculate the $dist_IP(q,A)$ we will use the trigonometric 
relation 
\begin{equation}
dist(q,A) = \frac{dist(q,C)}{\sin\alpha}
\end{equation}

The following is the crucial observation which underpins the whole approach.
\begin{itemize}
\item
In low dimensional space, $\sin\alpha \approx 1$. Thus $dist(q,A) \approx
dist(q,C)$  and thus the Angle Tree and kd-tree will behave in a similar
fashion.
\item
In high dimensional space (with low intrinsic dimensionality), $\sin\alpha
<< 1$, and thus $dist(q,A)$ will be potentially a much tighter lower bound than 
$dist(q,C).$ 
\item
In practice, real data sets rarely follow the equations of
a smooth manifold and there will be data points which lie outside the
manifold. Also, even if the data forms a smooth manifold, this manifold can be highly curved
making it difficult to approximate its local regions with an affine hyperplane. 
These two factors can cause data points to lie in the {\it error
region} shown in Figure \ref{fig:mainidea}. Then
the true nearest neighbors could be accidentally pruned.
\end{itemize}

\section{High Dimensional Behavior} \label{sec:highDimBehav}
We provide a rigorous justification for the behavior of the dihedral
angle ($\alpha$)  in high dimensional space as noted in Section \ref{sec:keyidea}.
Theorem 1 below is based on Figure \ref{fig:mainidea} but generalizes
to higher dimensions.

\begin{thm}
Given a random $d$-dimensional hyperplane ($IP$) in $\mathbb{R}^{D}$
and a random affine hyperplane of dimension D-1 (splitter) defined
by its normal vector $v\sim N(0,I_{D})$.  Let $\alpha$ be the dihedral
angle. If $D$ is large, $\sin\alpha$ converges 
to $\frac{\chi_{d}}{\sqrt{D-\frac{1}{2}}}$,
where $\chi_{d}$ is the Chi-distribution with $d$ degrees of freedom \cite{chiDistrib}.
If $d$ is also large, 
\begin{equation*}
\sin\alpha\sim N\left(\frac{\sqrt{d-\frac{1}{2}}}{\sqrt{D-\frac{1}{2}}},\frac{1}{2D-1}\right)\end{equation*} \label{thm:alphaDistribution} 
\end{thm}
\begin{IEEEproof} As $v$ is a D-dimensional random vector, we can express it
as $(x_{1},x_{2},\ldots,x_{D})$ and fix the d-dimensional IP as
the hyperplane spanned by the first d axes of $\mathbb{R}^{D}$. Here each
$x_{i} \sim N(0,1)$. Based on Figure \ref{fig:mainidea}, note $\alpha=\pi/2-\angle(v,\Pi v)$ 
where $\Pi v$ is the projection of $v$ onto $IP$. Thus $\Pi v=(x_{1},x_{2},..,x_{d},0,..,0)$.

\begin{equation*}
\sin\alpha=\cos\angle(v,\Pi v)=\frac{v\cdot\Pi v}{|v||\Pi v|}=\frac{\sum_{i=1}^{d} x_{1}^{2}}{\sqrt{\sum_{i=1}^{d} x_{1}^{2}}\sqrt{\sum_{i=1}^{D} x_{1}^{2}}} = \frac{\sqrt{\sum_{i=1}^{d} x_{1}^{2}}}{\sqrt{\sum_{i=1}^D x_{1}^{2}}}
\end{equation*}

The numerator and denominator are both chi-distributions
with $d$ and $D$ degrees of freedom respectively. For large $D,\,\chi_{D}\sim N(\sqrt{D-\frac{1}{2}},\frac{1}{2})$
\cite{fisher}. Since we assume $D$ to be very large, the denominator's variance
becomes insignificant relative to its mean, and so we can replace
it by its mean.
\begin{equation*} 
\sin\alpha\sim\frac{\chi_{d}}{\sqrt{D-\frac{1}{2}}}
\end{equation*}
 For the case when $d$ is also large, we apply \cite{fisher} 
to the numerator as well, yielding
\begin{equation*}
\sin\alpha\sim N\left(\frac{\sqrt{d-\frac{1}{2}}}{\sqrt{D-\frac{1}{2}}},\frac{1}{2D-1}\right)
\end{equation*}
\end{IEEEproof}

\section{Estimation of the Dihedral Angle} \label{sec:estimDihed}

\begin{figure}
\begin{centering}
\includegraphics[scale=0.55]{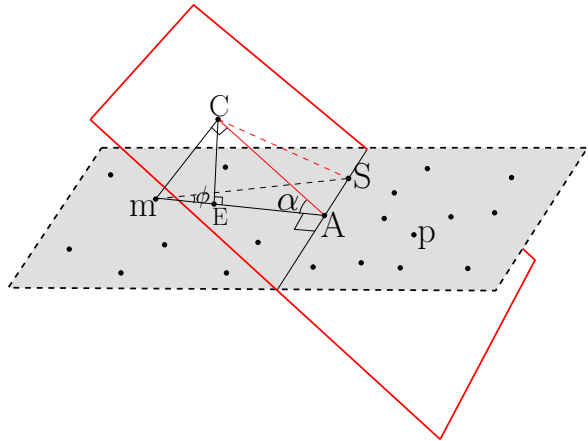} 
\par\end{centering}

\caption{$IP$ indicated by dashed line. Dots represent the data. $m$ is a
representative point of the data region. $p$ an arbitrary
point on the opposite side of the splitter (red hyperplane) to $m$.
We let this region have a LCD with $\epsilon=0$ for clarity.}

\centering{}\label{Flo:3dmainidea} 
\end{figure}
The dihedral angle $\alpha$ between two hyperplanes $P1$ and $P2$ of codimension 1 is 
related through their unit normals $n1$ and $n2$ by the equation \cite{dihed}
\begin{equation}
\cos\alpha = n1 \cdot n2  
\end{equation}
Thus if we know the two hyperplanes and their unique normals, calculating the dihedral angle is straightforward. 
The challenge in our case is that we do not know
the equation of the Intrinsic Plane (IP) and furthermore we assume that
the IP will change from region to region. In fact, by definition, manifolds are
locally like hyperplanes. We could use a PCA-like technique but that
would be computationally prohibitive as the complexity of PCA is $O(D^{3})$
where $D$ is the ambient dimension. We will propose a much simpler
and more efficient method which is also mathematically rigorous. We
refer the reader to Figure \ref{Flo:3dmainidea} for the discussion in
this section.

In Figure \ref{Flo:3dmainidea}, $m$ is the representative point and $\vec{mC}$ is
the normal from $m$ to the splitter. $\alpha = \angle CAm$ is the 
dihedral angle which can be obtained by projecting $\vec{mC}$ onto the IP.
Let $S$ be an arbitary point on the $IP$ (here it is shown to be on
the intersection of the two planes but it does not have to be). 

An important observation is captured in the following theorem. 
\begin{thm}\label{thm:cosses}
$\cos\angle CmS =\sin\alpha\cos\phi$ where $\alpha$ is the dihedral
angle and $\phi$ is the angle between $mA$ and $mS$ (see Figure \ref{Flo:3dmainidea}).
\end{thm}
\begin{IEEEproof} Since $\alpha$ is a dihedral angle, $\sin\alpha=\cos(\pi/2-\alpha)=\frac{\vec{mC}\cdot\vec{mA}}{|mC||mA|}$. Also
\[
\cos\angle CmS=\frac{\vec{mC}\cdot\vec{mS}}{|mC||mS|}
\]
 Substituting $\vec{mS}=\vec{mA}+\vec{AS}$ and $|mS|=|mA|/\cos\phi$
in the  above equation we get\\
 \begin{equation}\label{dihedBehav}
\cos\angle CmS=\cos\phi\frac{\vec{mC}\cdot\vec{mA}+\vec{mC}\cdot\vec{AS}}{|mC||mA|}\end{equation}
Since $\vec{AS}\bot\vec{mA}$, $\vec{AS}\bot\Pi(\vec{mC)}$ where $\Pi(\vec(mC))$
is the projection of $\vec{mC}$ onto the $IP$. Clearly also $\vec{AS}\bot\left(\vec{mC}-\Pi(\vec{mC)}\right)$,
since that is a vector normal to the $IP$. Then $\vec{AS}$ is normal
to the plane spanned by $\Pi(\vec{mC)}$ and $\left(\vec{mC}-\Pi(\vec{mC)}\right)$.
 Therefore $\vec{AS}\bot\vec{mC}$ and $\vec{mC}\cdot\vec{AS}=0$\\
 Then equation \ref{dihedBehav}  becomes 
\[
 \cos\angle CmS=\cos\phi\frac{\vec{mC}\cdot\vec{mA}}{|mC||mA|}=\cos\phi\sin\alpha
\]
\end{IEEEproof}

We should note that the proof of this theorem does not make use of
$S$ lying on the intersection between the $IP$ and the splitter,
and so this angular relation is true for any $\vec{mS}$ on the $IP$.

\begin{cor} $\forall \vec{v}\parallel IP\,\,\,\, \angle(\vec{v},\vec{mC})\le\pi/2-\alpha$
\comment{No vector that lies on the
$IP$ can make an angle greater than $\pi/2-\alpha$ with $\vec{mC}$}
\end{cor}
\begin{IEEEproof}
When $\pi/2\ge\phi\ge0$, $\cos\angle CmS\le\sin\alpha=\cos(\pi/2-\alpha)$.
As $\phi\rightarrow0$, $\cos\angle CmS\rightarrow\cos(\pi/2-\alpha)$
and hence $\angle CmS\rightarrow\pi/2-\alpha$ from below. 
\end{IEEEproof}

\begin{cor} The dihedral angle $\alpha\ge\angle mSC$
\end{cor}
\begin{IEEEproof}
Since $\angle CmS\ge\pi/2-\alpha$, $\alpha\ge\pi/2-\angle CmS=\angle mSC$
\end{IEEEproof}

Theorem \ref{thm:cosses} and the corollaries suggest the following algorithm for estimating alpha:

We take the center of the data region and sample a constant number
($k)$ of random data points within the region. We then subtract the
center from all of these points to obtain $k$ random vectors that
will represent the region (there are many other methods for obtaining 
these $k$ random vectors). We calculate the angle that all of these
vectors make with the normal to the splitter $\vec{mC}$ (each region
has one splitter). The smallest of these $k$ angles will be within
a small range of $\pi/2-\alpha$ ($\angle CmA$ in Figure \ref{Flo:3dmainidea}),
the true angle between $\vec{mC}$ and the $IP$ (See the Appendix for
pseudocode of this procedure). In order for this to work, at least one 
of the random vectors must be within a small angle of $\vec{mA}$ (as 
close to parallel as possible). The probability of this occurring 
depends on the dimensionality of $IP$, as we will show.

\subsection{Evaluation of Random Vector Strategy}

\begin{figure}
\begin{centering}
\includegraphics[clip,scale=0.50]{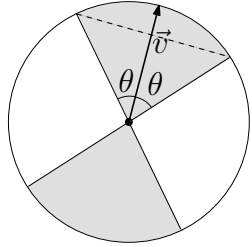} 
\par\end{centering}
\caption{$\theta$ determines the width of the target region (shaded). If a random vector 
makes an angle close to $180^o$ to $\vec{v}$, we simply multiply it by -1 
in order to obtain a vector close to $\vec{v}$. Hence our target region's volume is 
doubled. The dashed line divides the target region into two parts; a hypercone 
and a hypersphere cap.}
\label{probMissCone}
\end{figure}

In order to evaluate this strategy, we form the following problem.
We assume the data is evenly distributed within a d-ball of unit radius,
centered at the origin, and we have some fixed vector $\vec{v}$
within this space. We then generate $k$ random vectors within the
ball, and calculate the probability that not one of these is
within some small angle $\theta$ (error angle) of $\vec{v}$ (please refer 
to Figure \ref{probMissCone} for illustration). We
denote the volume of the d-ball segment spanning the vectors that
are close to $\vec{v}$ by $s$, and the volume of the entire d-ball
by $S$. 
Since we assumed the data to be evenly distributed, every random
vector has a constant probability $s/S$ of landing within the aforementioned
segment. $s$ has volume given by the formula for the d-dimensional cone $Base \times Height/d$ \cite{volCone} plus
the volume of the hypersphere cap given by the formula described in \cite{sphereCap}.
Combining these two formulas, the ratio $s/S$ has the value given by:
\begin{equation} 
\label{eq:sonS}
s/S = 2\left(\frac{1}{2}-\cos\theta\frac{\Gamma[1+\frac{d}{2}]}{\sqrt{\pi}\Gamma[\frac{d+1}{2}]}{}_{2}F_{1}\left(\tfrac{1}{2},\tfrac{1-d}{2};\tfrac{3}{2};\cos^{2}\theta\right)\right)+\frac{2\cos\theta\sin^{d-1}\theta\cdot\Gamma[1+\frac{d}{2}]}{d{\sqrt{\pi}\cdot\Gamma[1+\frac{d-1}{2}]}}
\end{equation}
where $\Gamma$ is the Gamma function and $_{2}F_{1}$ is Gauss' hypergeometric function \cite{gaussFunc}. Note 
that this ratio depends only on $\theta$ and the intrinsic dimension $d$. Monte Carlo 
experiments confirm the values given by this formula.

The probability that all $k$ vectors miss the segment is
then $(1-s/S)^{k}$. We show that this probability can be made very
small for $k \thickapprox 2,000$, $\theta\thickapprox30^{o}$ while
$d\leq10.$ Beyond that dimension, to maintain a small $\theta$,
$k$ would have to grow exponentially. So in Section \ref{sec:estimDihed}, 
as long as $d$ is not too large, we can get a vector $\vec{mS}$ 
that is within $\theta$ of $\vec{mC}$.
\begin{figure}
\begin{centering}
\includegraphics[scale=0.28]{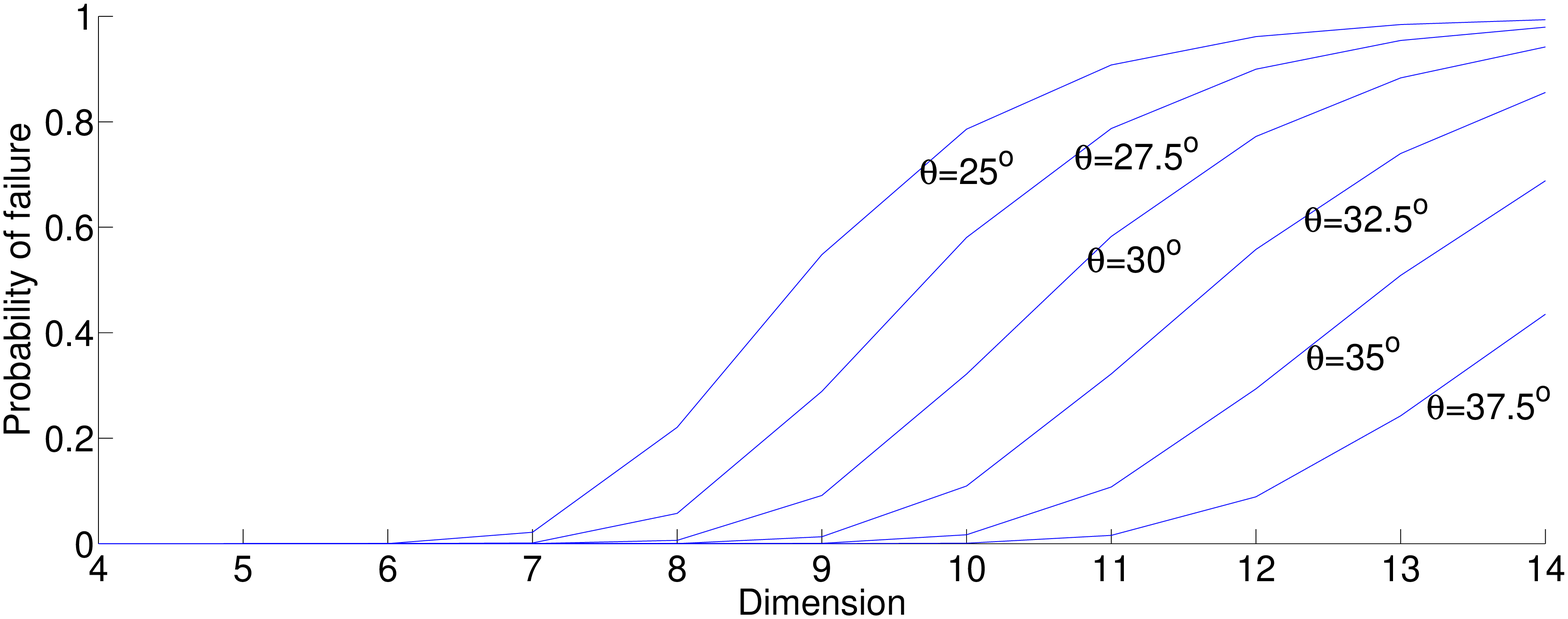} 
\par\end{centering}

\caption{Probability of all the 2000 random vectors missing a segment (failure) of width $\theta$ within a d-dimensional 
ball is sensitive to both $\theta$ and $d$. The formula used to build this graph is $(1-s/S)^{2000}$ where s/S is given by Equation \ref{eq:sonS}.}

\label{Flo:probmis} 
\end{figure}

One obvious problem with this approach is that if the data is noisy,
then the vectors we generate do not lie exactly on the $IP$. This
causes some inaccuracy in determining $\alpha$. Our approach for
dealing with this involves simply ignoring a constant portion of the
most extreme angles, attributing them to noise. This is described
further in Section \ref{sec:exp}.

Another problem occurs when the region in question is large, where
the manifold curves sharply with respect to the splitter. In this case
there is no low dimensional $IP$ that approximates the data well.%
\begin{figure}
\begin{centering}
\includegraphics[scale=0.35]{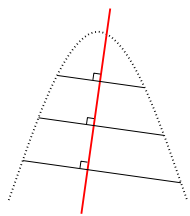} 
\par\end{centering}

\caption{One of the random vectors needs to be (close to) normal to the splitter
for the Angle Tree to behave like a regular kd-tree. Here the manifold despite being 1-dimensional, bends 
sharply w.r.t. the splitter, and so there are many such vectors. When the region is made smaller by further splits, 
the manifold will become {\it flatter}, and so the directions of the vectors generated in the region will be more limited.}

\label{Flo:bigRegion}
\end{figure}

The case in Figure \ref{Flo:bigRegion} is not a pathological one.
Despite the data having intrinsic dimension of 1, the region in question
is too large to be approximated by a 1-dimensional $IP$. For this
reason, we consider this data region to be 2-dimensional for our purposes.

In order for this technique to not prune any regions incorrectly,
we must find that $\alpha=\pi/2$ here - we use the regular kd-tree
search lower bound, since $dist(q,A)=dist(q,C)/\sin\alpha=dist(q,C)$ in this case.
In order for this to happen, at least one of the randomly generated
vectors must make an angle of $\pi/2$ (or close to it) with the splitter.

Then, we do not gain or lose anything compared to the standard kd-tree
pruning method, in this case. As we continue to split the data, the
regions will become small enough that they can be approximated by
a low dimensional $IP$.

Another possibility: if the data lies on a d-dimensional manifold,
but we consider a region where the manifold curves sharply, but only
in one other dimension. In this case, the region can still be approximated
by a d+1 dimensional $IP$. For example, the data in Figure \ref{Flo:bigRegion}
could have a very large ambient dimension, but this region can still
be approximated by a 2-dimensional $IP$. Our method will be robust
to this possibility, since from Theorem \ref{thm:alphaDistribution},
the angle between the d+1 dimensional $IP$ and the D-1 dimensional
splitter is still very likely to be much smaller than $\pi/2$, giving
the angle tree a large performance boost over the standard kd-tree
pruning.

\subsection{Effect of Error Angle on Pruning Calculation}

We refer again to Figure \ref{Flo:3dmainidea}. Let $\vec{mS}$ be
an extension of the randomly generated vector making the smallest
angle with $\vec{mC}$. Then if we have some estimation of $d$, and
by referring to Figure \ref{Flo:probmis}, we can say that with high
probability $\vec{mS}$ is within $\theta$ (of some appropriate size)
of $\vec{mC}$ -- $\phi<\theta$. Then making use of Theorem \ref{thm:cosses}
again, $\cos\angle CmS=\sin\alpha\cos\phi>\sin\alpha\cos\theta$,
and from this we can derive \[
|mA|>\cos\theta|mC|/\sin\angle CSm\]
 where $\angle CSm$ is our slightly erroneous estimate for $\alpha$
and $\cos\theta$ is compensation for this error.

Putting the query point $q$ in the place of $m$, we now have a 'safe' lower bound on $dist(q,A)$, 
and it tightens as $\theta$ is made smaller (by generating more random 
vectors, for instance). This bound should still be much tighter than $dist(q,C)$.

\section{Error Region} \label{sec:errorReg}

\begin{figure}
\begin{centering}
\includegraphics[scale=0.55]{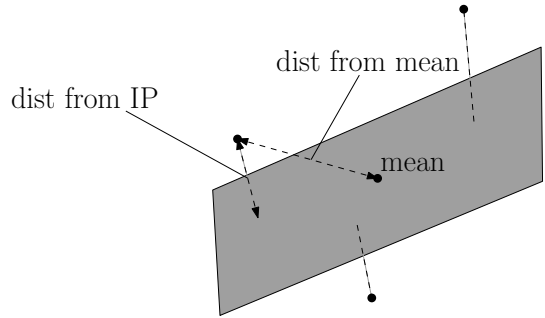} 
\par\end{centering}
\caption{Illustration of LCD. If $\epsilon$ is small, then the distance of data point $p$ from $IP$
is a relatively small component (on average over all $p$) of the distance of $p$ from the mean of the data region.
}
\label{fig:cov-dim} 
\end{figure}

\begin{figure}
\begin{centering}
\includegraphics[scale=0.55]{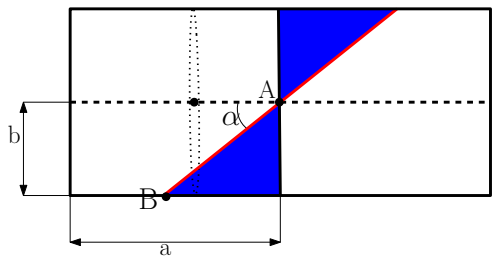} 
\par\end{centering}

\caption{A is the geometric mean of the data. Error Region is
blue. Red line is splitter. Vertical black line through A is perpendicular
(ideal) splitter. The hypercylinder is made up of the $IP$ (a d-ball with 
radius $a$ - represented by the horizontal dashed line) and an infinite number 
of (D-d)-balls centered on each point in the $IP$ (not data point) and perpendicular to the $IP$. 
In the case when $d=1$ and $D=2$, the hypercylinder is a rectangle as shown. 
Data is evenly distributed within the hypercylinder.}

\label{Flo:errorRegion} 
\end{figure}

In Figure \ref{fig:mainidea} we noted an error region, where any
point $p$ falling into it has the property $dist_{IP}(q,p)<dist(q,A)$
and hence breaks one of our assumptions. The cell containing this
point might be pruned using our algorithm, even though it contains
a neighbor that is nearer to $q$ than the best-so-far found neighbor.
In order to analyze this possibility we need a model of the region
in question. For this, we make use of the property of LCD as described
in \cite{rptreeLect}; if the data has covariance dimension $(d,\epsilon,r)$ for some $r$, 
then any data region that fits entirely inside a ball of radius $r$ will have 
\begin{equation*}
(avg\, dist^{2}\, from\, IP)\leq\epsilon\cdot(avg\, dist^{2}from\, mean)
\end{equation*}
where the $avg\, dist$ and $mean$ (geometric mean on all the dimensions) 
are taken across all the data points in the region (see Figure \ref{fig:cov-dim}).

We assume that $\epsilon$ is small ($0\le\epsilon\le0.1$). If we
let the noise be uniform, we have the following model: points are
uniformly distributed within the hypercylinder shown in Figure \ref{Flo:errorRegion}.
The $IP$ itself is a $d$-ball with radius $a$. Centered on each
point on the $IP$ we have a $(D-d)$-ball normal to $IP$ with radius
$b$. $a$ and $b$ are chosen so that the data being distributed
uniformly within the hypercylinder, has $avg\, distance{}^{2}$ within
$IP$ equal to 1, and the $avg\, distance{}^{2}$ from subspace equal
to $\epsilon$. The $avg\, distance{}^{2}$ from mean is then equal
to $1+\epsilon$, so the model satisfies the above constraint (since
$\epsilon\le\epsilon+\epsilon^{2}$) while being close to the maximal
noise case (since $\epsilon^{2}$ is insignificant for small
$\epsilon$). Then the ratio of the volume of the error region $v$ to the volume
of the entire region $V$ described by the model corresponds to the
proportion of points that fall within the error region. 

We obtain it by integrating the Error Region volume along $z$ 
(any axis within $IP$) from $B~(z=-b/\tan\alpha)$ til $A~(z=0)$ 
and multiplying by 2. We denote the volume of a $k$-ball with radius 
$l$ by $B^{k}(l)$ and the volume of a hypersphere cap of height 
$m$ and dimension $k$ by $C_{l}^{k}(m)$. By considering the intersection of the $IP$
(which we assumed to be a d-ball) and a 2-dimensional plane through $A$, and noting that this 
intersection must be a circle, we see that the radius of the (d-1)-ball in the $IP$ subtended
by the integration slice $z$ units away from $A$ will have radius $\sqrt{a^{2}-z^{2}}$. 
The volume subtended from the (D-d)-ball in the noisy directions will 
equal $C_{b}^{D-d}\left(b+ztan\alpha\right)$. Hence the total
volume of the Error Region is given by:

\begin{equation}
\frac{v}{V}=\frac{2\int_{-b/\tan\alpha}^{0}\, B^{d-1}\left(\sqrt{a^{2}-z^{2}}\right)\times C_{b}^{D-d}\left(b+ztan\alpha\right)dz}{B^{d}(a)\times B^{D-d}(b)}\end{equation}

Where $D$ is the ambient dimension, $d$ is the intrinsic dimension
of $IP$, $a=\sqrt{3/d}$, $b=\sqrt{3\epsilon/(D-d)}$,
and $\epsilon$ is the LCD coefficient. We often find that this ratio
is very small when $d\ll D$ and $\epsilon\le0.1.$\\

\begin{prop}
When $a=\sqrt{3/d}$ and $b=\sqrt{3\epsilon/(D-d)}$, then the
data distributed uniformly within the hypercylinder, has $avg\, distance{}^{2}$
within $IP$ equal to 1, and the $avg\, distance{}^{2}$ from subspace
equal to $\epsilon$.
\end{prop}
\begin{IEEEproof}
We assumed that the first d axes are: $x_{i}\sim[-a,a]$. Then
\begin{equation*}
E\left(\sum_{i=1}^{d} x_{i}^{2}\right)=\sum_{i=1}^{d} E\left(x_{i}^{2}\right)=d\cdot E\left(x_{1}^{2}\right)=1\Rightarrow E\left(x_{1}^{2}\right)=\frac{1}{d}
\end{equation*}
\noindent \begin{flushleft}
because of the linearity of expectation, and all the $E(x_{i}^{2})$'s are equal. \\
$E\left(x_{i}^{2}\right)=\frac{\int_{-a}^{a}\, x^{2}}{2a}=\frac{\frac{1}{3}\times2a^{2}}{2a}=\frac{a^{2}}{3}=\frac{1}{d}$

giving $a=\sqrt{3/d}$ \\
By a similar process we get $b=\sqrt{\frac{3\epsilon}{D-d}}$\\
\par\end{flushleft} \end{IEEEproof}
It can be seen that the Error Region is smaller if $\alpha$ is made
larger (our splitter is closer to the ideal splitter). 
In Section \ref{sec:disc} we give one possible way to achieve this.

\section{Experiments and Analysis} \label{sec:exp}

\subsection{Design of Experiments}

In this section the Angle Tree is experimentally compared against
the cover tree and locality sensitive hashing (LSH). They are the
two most popular and most recent developments in high-dimensional
NNS. The Angle Tree is also compared against Partial Brute Force (PBF).
If for an experiment the Angle Tree achieved an average accuracy of
90\% or 0.9 when searching for k-nearest neighbors (meaning that all
k neighbors returned were the true nearest neighbors 90\% of the time, as as verified
by a full brute force search), then PBF, searching randomly through
unindexed data, must search through $\%(100\times0.9^{1/k})$ to achieve
the same average accuracy.

The comparison will be based on preprocessing time, space complexity,
query time and accuracy. The comparison vis-a-vis kd-trees and rp-trees
is not reported as the NNS query using the latter two data structures
degenerates to a brute force search in high dimensions if the standard
pruning bound is used.

It it worth recalling that the original rp-tree (as proposed by the
authors) can only be used to efficiently answer \textit{near neighbor
search }\textit{\emph{(where we only search through a single cell
and terminate - see Section \ref{sec:keyidea} for details of why this often misses
the true nearest neighbors).}} If the kd-tree search strategy is used
with the rp-tree, then the search degenerates to brute force. In fact,
the principal aim of Angle Trees is to extend rp-trees (and kd-trees)
to make NNS in high dimensions efficient and accurate.

\subsubsection{Implementation Details}

The Angle Tree was implemented on top of the rp-tree code base available
from the authors website \cite{Dasgupta}. We did not use the error
angle adjustment in our implementation, since speed seemed to be more
of a bottleneck than accuracy. All experiments were performed on an
AMD dual core 1 GHz processor with 4GB of RAM. In our experiments
we will refer to the Angle Tree as angle rp-tree to emphasize its
connection with random projections and rp-trees.

Cover tree code was obtained from the authors website \cite{Beygel2006}.
This code was used to attempt to preprocess all of the data sets used.
The cover tree performance data is taken from \cite{Comparison2005}.

LSH (where the hash functions are a series of random projections)
performance was inferred from the performance of \textit{near neighbor
search }\textit{\emph{in a single rp-tree. }}If $t$ independent rp-trees
are built, then each one has a similar probability $p$ of having
both the query point $q$ and its nearest neighbor hashed to the same
cell. Then the probability of not finding the nearest neighbor in
any of the $t$ trees is $(1-p)^{t}$. The average search time is
then $xt$, where $x$ is the average search time for a single rp-tree
near neighbor search. The probability $p$ obviously depends on the
size of the cells - or the number of LSH hash buckets. This data is presented
in Table \ref{Flo:LSHresults}.

\subsubsection{Data Sets}

Several well known real data sets were used for comparison. We also
generated two synthetic data sets on high-dimensional spheres. The
details are listed in Table \ref{tab:data} in the Appendix. It is
worth emphasizing that we used extremely high-dimensional, real world
data sets (e.g., Reuters Bag of Words, Mnist and Yale Face image database)
which are extremely noisy and hard to index. For overview of all the
data sets used, see Appendix.

\subsubsection{Comparison Metric}

The fundamental metric used to compare the data structures is \textit{number
of distance calculations (NDC)} - which is defined as the number of
times euclidean distance between two points is calculated. This metric
is hardware independent. 

NDC is used as a proxy to measure the running time of the algorithm,
being by far the most computationally intensive part of the algorithm.
NDC correlates very strongly with the actual running time of the algorithm,
and was used in \cite{Comparison2005} to evaluate the cover tree. 

\subsubsection{Parameters}

There is one important parameter in angle rp-trees whose effect needs
to be measured, namely, the \textit{Ignore Outlier (IOut)} parameter.

IOut controls the effect of noise in the data during the estimation
of the dihedral angle. IOut is the proportion of the angles between
the random vector and the normal to the splitter ($\vec{mC}$ in Figure
\ref{Flo:3dmainidea}) that will be ignored. The assumption is that
very small angles are caused due to the presence of outliers, since
outliers do not lie on the $IP$ and do not conform to Theorem \ref{thm:cosses}.

For example, suppose one thousand random data vectors are generated
and for each such vector, the angle with the normal to the splitter
is calculated. These angles are then sorted in an increasing order
and if the IOut value is 0.1, then the $100^{th}$ angle is the estimated
value of the dihedral angle.

IOut controls the trade-off between accuracy and search time. High
values of IOut will inflate the estimation of $dist_{IP}(q,A)$.
This will result in more aggressive pruning but could result in some
nearest neighbors being missed.

Another parameter that we vary is the number of levels in the tree.
This is significant for the LSH analysis as it determines the number
of hash buckets. More and smaller buckets usually translates to lower
accuracy but faster search time.

The other LSH parameter is $t$, the number of trees (hash functions)
over which we infer the performance of LSH.

\subsubsection{Experiments}

The following three experiments were conducted. In E1, we measure
preprocessing efficiency, and in E2 and E3 we measure search efficiency. 
\begin{lyxlist}{00.00.0000}
\item [{{E1}}] For each data set, the angle rp-tree was constructed and
the NDC was recorded. We note that NDC has the same complexity as
angle computations and projection onto splitter (O(D), since they
all involve a dot product) and so for the angle rp-tree we include
them in the NDC value. We also used the cover tree code to see whether
preprocessing the various data sets would cause a crash. We used the
standard rp-tree (with various number of levels) as an implementation
for LSH. 
\item [{{E2}}] For each data set, one thousand random data points were chosen
and used to simulate queries. In this way the query points were 
guaranteed to also come from the data manifold. The angle rp-tree was then used
to search for 1-NN with various IOut parameter values for each query point. The NDC was
recorded during the search process and then averaged over the one
thousand queries. 
\item [{{E3}}] For the Mnist and KDD data, a search for 2-NN was also
carried out as comparative data was available from \cite{Comparison2005}. 
\end{lyxlist}

\noindent %
\begin{figure*}
\noindent \begin{centering}
\includegraphics[bb=0bp 180bp 603bp 680bp,clip,scale=0.40]{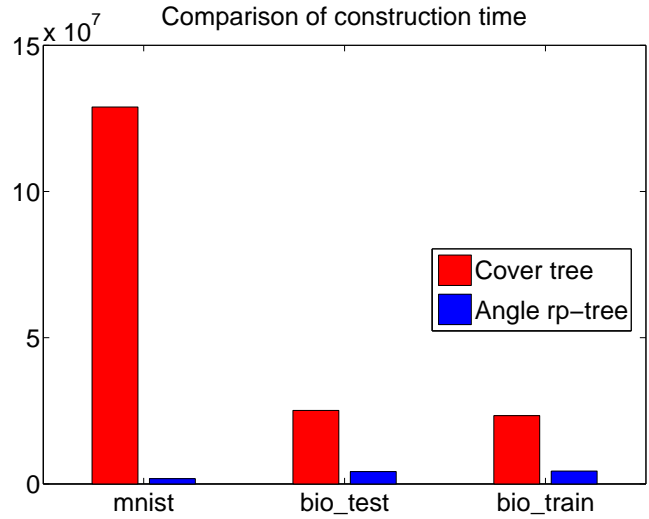} 
\par\end{centering}
\caption{Here is the main advantage of the angle rp-tree over the cover tree;
cover tree preprocessing time - the number of distance computations
in the construction of the data structure - is exponential in $d$.
{[}11{]} }

\label{fig:preprocess} 
\end{figure*}

\begin{figure*}
\begin{centering}
\includegraphics[trim=130bp 20bp 120bp 20bp,clip,scale=0.40]{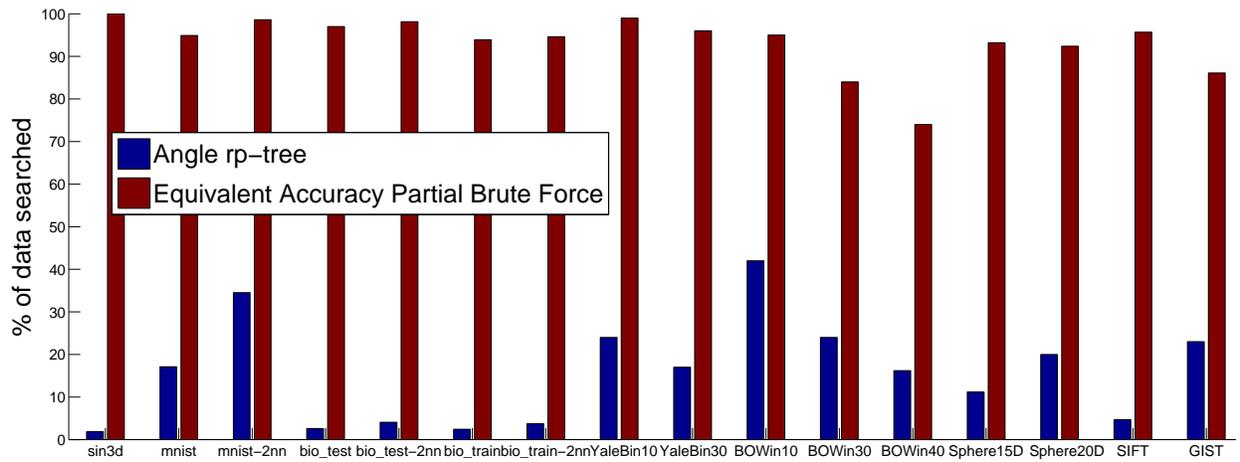}
\par\end{centering}

\caption{Speedup over partial brute force. The amount of data that the brute
force randomly searches through in order to get the same average accuracy
as our method. To get $x\%$ accuracy, partial brute force must on
average search through $x\%$ of data for 1-NNS and $10\sqrt{x}\%$
of data for 2-NNS. 'in10' means that the IOut value for an experiment
was 0.1 -- we ignored 10\% of the most extreme angle values.}

\label{Flo:angletreeVSpbf}
\end{figure*}

\begin{figure*}
\begin{centering}
\includegraphics[scale=0.33]{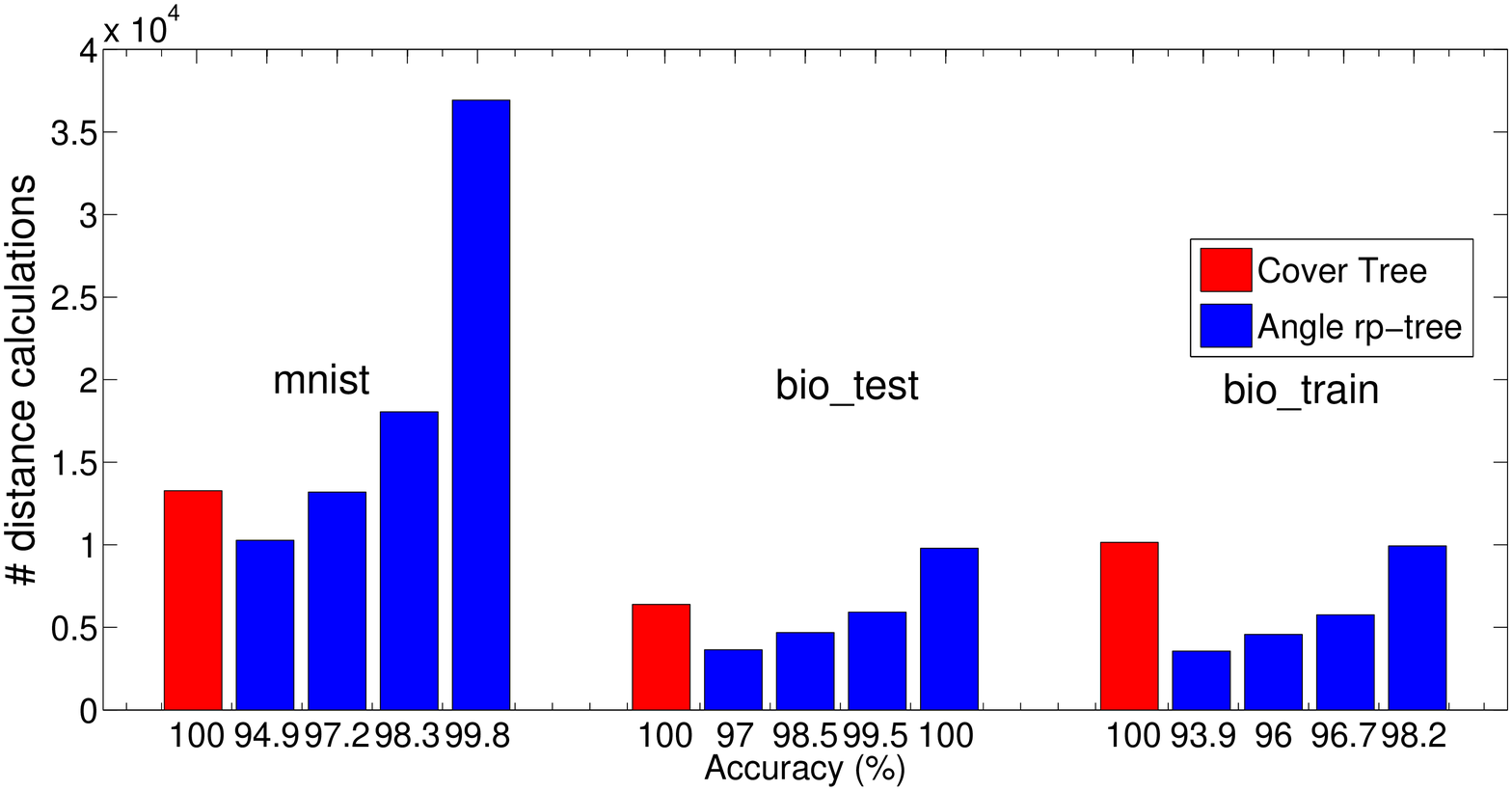} 
\par\end{centering}

\caption{Search performance of angle rp-tree vs cover tree for 1-NNS. Various
IOut values are used to balance speed and accuracy. While cover tree
is capable of 100\% accuracy, our method can come very close with
comparable running time.}

\label{Flo:cover1NNS} 
\end{figure*}

\begin{figure*}
\begin{centering}
\includegraphics [scale=0.33]{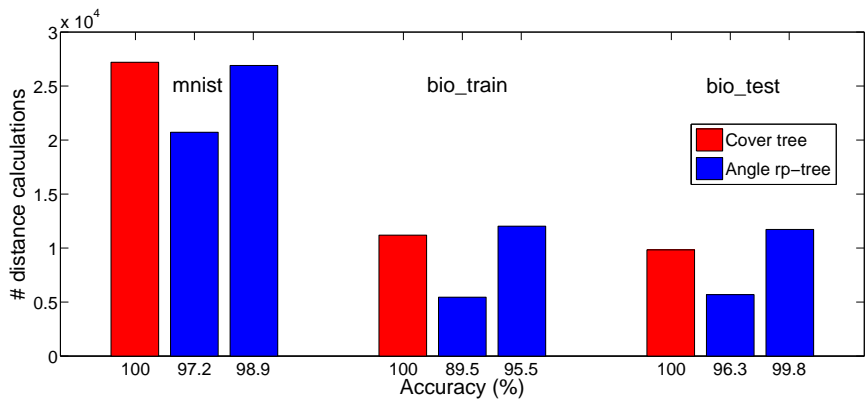} 
\par\end{centering}

\raggedright{}\caption{Search performance of angle rp-tree vs cover tree for 2-NNS. Various
IOut values are used to balance speed and accuracy. While cover tree
is capable of 100\% accuracy, our method can come very close with
comparable running time.}

\label{Flo:cover2NN}
\end{figure*}

\subsection{Results and Analysis}

We now report on the results of the three different experiments under
varying conditions.

\subsubsection{E1}

\paragraph*{Cover Tree}

See Figure \ref{fig:preprocess}. Theoretically the preprocessing
time of angle rp-trees is nearly of the order of a standard kd-tree
or rp-tree, which is $O(nlogn)$. The only additional overhead is the
calculation and storage of the dihedral angle for each data region. Since the
number of data regions (nodes) is $O(n)$, and we calculate a constant
number of angles ($k$) in each data region, the complexity of angle
rp-tree is $O(nlogn+nk)$ which is $O(nlogn)$.

For cover trees, the theoretical preprocessing time is:\\
 $O(c^{6}n\log n)$ where $c$ (the KR-dimension) $\sim2^{d}$
and potentially much larger for noisy data. A single noisy point can
make $c$ grow arbitrarily \cite{Karger2002}.

The preprocessing in our experiments for the cover tree and angle
rp-tree are shown in Figure \ref{fig:preprocess}. It is clear that
the preprocessing time of the angle rp-tree is a small fraction (less
than 1\%) of the cover tree.

Another factor that is not reflected in the results is that even though
the cover tree space requirements are O(n) just like the kd-tree,
the data points appear multiple times in the structure. Thus cover
tree's space complexity has a higher constant. Its memory usage is
often three to five times greater than the size of the data set \cite{Comparison2005}.

We have only shown results on three data sets for which the comparison
data was available. The cover tree crashed during construction for
the Reuters Bag of Words and the synthetic sphere databases. It crashed
likely due to overflow of the recursion stack during construction \cite{Shi06}
due to those data sets having an intrinsic dimension that is too large.

\paragraph*{LSH}

See Table \ref{Flo:LSHresults}. The preprocessing time of LSH is
similar to rp-trees except that there are $t$ trees (hash functions)
and so roughly $t$ times as much preprocessing (the constant number
of angle calculations that are not required in LSH become relatively
trivial anyway, when the data set is large). However, LSH preprocessing
time has no dependence on $d$ and is unaffected by noise, and so
is fairly efficient.

Due to there being $t$ trees and each data point is \emph{hashed}
into each tree, LSH space complexity is significantly greater than
the angle rp-tree. It is difficult to say how much more exactly since
only the hash values of the data are stored, but \cite{multi-probe,Joly2008}
reports it to be significant - several times the size of the original
data. As we will see later, when $t$ is reduced, then accuracy and/or
search speed suffers. For larger data sets than the ones tested here,
\cite{multi-probe,Joly2008} note that LSH often requires $t$ to
be of the order of 100 to obtain acceptable search speed and accuracy.

\subsubsection{E2}

\paragraph*{PBF}

See Figure \ref{Flo:angletreeVSpbf}. Here we see the speedup of angle
rp-tree over PBF. Even in 30,000+ dimensional data like the images
database, we find the true nearest neighbors 95\%+ of the time while
searching <20\% of the data. This indicates very strongly that the
curse of dimensionality can be managed when data has a low intrinsic
dimensionality, without any dependence on the ambient dimension. We
believe that if the database had more items (Yale Face has <2500 items)
the angle rp-tree indexing would give an even greater speedup, similar
to the KDD data sets.

The noisiest (or least structured) and hence most difficult data set
was the Reuters Bag of Words data set. For this data set we had to
search through 42\% of the data in order to achieve 95\% accuracy
for 1-NNS. We could also search through 24\% of data to achieve 85\%
accuracy. This second result seems to be more practical to us, though
the user would have to accept a 15\% error rate. Additionally, in
the 15\% of cases where an incorrect neighbor was returned, it was
usually a small error. As far as we know, no one else had indexed
this kind of data set with any significant success.

The sin3D, as well as the 15D and 20D Sphere synthetic data sets are included to indicate
what sort of complexity of structure the real world data sets must
have, if they have similar performance to the synthetic data sets
whose structure is known. The 15D Sphere data set gave much better
results, and was close in performance to the KDD bio data sets, whereas
the 20D sphere data set was closer to the Reuters Bag of Words data
set. This is consistent with our analysis in Section \ref{sec:estimDihed}, which suggested
that the intrinsic dimensionality must be not much greater than ten
in order for the dihedral angle to be accurately estimated. 

Results for the sin3D data set were 100\% accurate, with most of 
the angles being over $80^o$. The angle rp-tree reduced to 
the kd-tree for this low-dimensional data set.

\paragraph*{Cover Tree}

See Figure \ref{Flo:cover1NNS}. The angle rp-tree achieves accuracy
well over 95\% with search speed similar to or faster than the cover
tree, although the cover tree achieves 100\% accuracy.

The angle rp-tree seems to slow down significantly as we try to approach
100\% accuracy with noisy data. We search through twice as much data
for the mnist data set in order to increase accuracy from 98.3\% to
99.8\%. This is likely because the $IP$ starts to include more and
more rare, noisy directions as we reduce the IOut parameter value.
As the $IP$ grows in dimension, the dihedral angle grows quite rapidly
causing the pruning multiplier $1/\sin\alpha$ to shrink. This causes
the algorithm to prune a lot less cells for very little gain in accuracy
(when the noisy points turn out to be the true nearest neighbors).
Due to this we can only really achieve 100\% accuracy as well as fast
running time when the data is not noisy, and hence can be well estimated
by a low dimensional $IP$. For noisy real world data, we must at
this stage settle for 95+\% accuracy.

\paragraph*{LSH}

See Table \ref{Flo:LSHresults}. In this table $t$ and $p$ are chosen
so that the accuracy corresponds to one of the results in Figure \ref{Flo:angletreeVSpbf}.

We can see that for the Reuters Bag of Words and mnist data sets,
LSH with large $t$ parameter is often several times faster on average
than the angle rp- tree, while having a similar accuracy. However,
when we reduce $t$ (in order to alleviate the space usage) but wish
to maintain accuracy, a tree with less levels must be built. Then,
since the leaf cells will contain more data, the probability that
the leaf cell into which the query point is hashed will contain its
true nearest neighbor, is made higher. This in turn makes LSH slower.
In fact, it can be seen that when $t$ is made too small, LSH becomes
significantly slower \emph{and} less accurate than the angle rp-tree.

It is also worth noting that in Table \ref{Flo:LSHresults}, when
considering the Yale Face image database, LSH is significantly slower
and less accurate (as well as more space consuming) than the angle
rp-tree. We believe this is because the Yale Face image database is
less noisy than mnist and Reuters BOW, and can be better approximated
by a low dimensional $IP$.

\subsubsection{E3 }

\paragraph*{PBF}

See Figure \ref{Flo:angletreeVSpbf} - those columns labeled 2nn.
The angle rp-tree searches through approximately twice as much data
on average for 2-NNS than 1-NNS. This is logical since the second
nearest neighbor is further away from the query point, making the
search sphere (see Section \ref{sec:keyidea}) larger, causing the algorithm to prune
less subtrees. In this way the angle rp-tree behaves as it should.
The kd-tree operating in the intrinsic space of the data would likely
perform in a similar way.

\paragraph*{Cover Tree }

See Figure \ref{Flo:cover2NN}. These results are roughly the same
as in E2. The angle rp-tree introduces a small probability of error
while achieving comparable search speed to the cover tree.

\section{Discussion} \label{sec:disc}

\paragraph*{Multi-probe Locality Sensitive Hashing}

This is a variation of LSH, where instead of only checking one bucket,
multiple buckets are probed. In some implementations, the buckets
are sorted by probability that they contain the true nearest neighbor,
and only the top few are checked. The purpose of this is to reduce
the space requirement of LSH by reducing the number of hash functions
that are needed to maintain high accuracy.

The problem with this approach is that a constant number of buckets
will be checked with each query, whereas with the angle rp-tree it is
often the case that the entire tree may be pruned after checking one
or two cells. There is no set number of buckets to check.

Additionally, current implementations of multi-probe LSH are based
on calculating the distance of the query point to the splitter \cite{multi-probe}.
However, as we saw in Theorem \ref{thm:alphaDistribution}, since
random splitters make significantly varying angles with the data,
this distance alone is not the best way of ranking the cells. The
angle rp-ree seems to be a promising way to improve current multi-probe
LSH implementations.%

We have shown that even at this early stage, our algorithm can compete
with the state of the art algorithms in terms of running time and
accuracy, while being significantly superior in terms of space and
preprocessing requirements. Additionally, it is likely to be much
simpler to use and implement, since it is not a major modification
to the well known kd-tree. We have also provided some new insight
on the nature of the curse of dimensionality in the context of NNS.

\paragraph*{Future Work}

A simple improvement to the random splitter (as used in the rp-tree)
would be a splitter whose normal vector lies on or very close to the
$IP$, but is still random within this restricted space. This can
be achieved by generating some constant number of random vectors in
the data as before, and then averaging them, finding their median,
or combining them in some arbitrary way. A splitter generated thus
would make a much larger angle with $IP$ than a random splitter,
and make the error region discussed earlier much smaller.

It seems very likely to us that there is a superior and more robust
method for estimating the dihedral angle with noisy data, such as
a Kalman filter type of process.

The way we deal with noise in the data, and the effect it has on the
dihedral angle estimation, is quite simplistic. It is almost certain
that there are more robust ways for doing this.

The exact relationship between the error angle compensation (multiplying
by $\cos\theta$) and the search speed hasn't been analyzed yet. In
fact, no guarantees on search speed, even in the average case, have
been established.

\appendix

\vspace{3.2cm}
\noindent \begin{flushleft} \begin{table*}
\begin{centering}
\begin{tabular}{|>{\raggedright}p{1.1in}|r|r|r|>{\centering}p{1.3in}|}
\hline 
Data set  & Size  & D  & d  & Description\tabularnewline
\hline
\hline 
sin3D  & 10,000  & 3  & 3  & rp-tree example data\tabularnewline
\hline 
mnist \cite{mnist}  & 60,000  & 784  & 5-10  & Handwriting\tabularnewline
\hline 
bio\_train \cite{kdd}  & 145,751  & 75  & 6  & KDD\tabularnewline
\hline 
bio\_test \cite{kdd}  & 139,658  & 74  & 6  & KDD\tabularnewline
\hline 
Extended Yale Face B Cropped \cite{yale1,yale2}  & 2432  & 32,256  & 8  & Human faces. Various people. Varying lighting\tabularnewline
\hline 
Reuters Bag of Words (BOW) \cite{reutersBOW}  & 11,887  & 6100  & 10-40  & News articles
word counts\tabularnewline
\hline 
15D Sphere  & 100,000  & 15  & 14  & Synthetic\tabularnewline
\hline 
20D Sphere  & 100,000  & 20  & 19  & Synthetic\tabularnewline
\hline 
SIFT\cite{sift} & 1,000,000 & 128 & 15-30 & Image descriptors\tabularnewline
\hline
\end{tabular}
\par\end{centering}
\noindent ~\\
\caption{Overview of the data sets. \textbf{D} is the dimensionality of the
data set and \textbf{d} is the approximate intrinsic dimension.}

\label{tab:data} 
\end{table*}

\par\end{flushleft}

\begin{center}
\begin{table*}
\begin{centering}
\begin{tabular}{>{\raggedright}p{0.5in}|>{\raggedleft}p{0.5in}|r|>{\raggedleft}p{0.5in}|>{\raggedleft}p{0.9in}|>{\raggedleft}p{0.9in}||>{\raggedleft}p{0.5in}|r}
\hline 
\multicolumn{1}{|>{\raggedright}p{0.5in}|}{Levels in tree } & Avg. Per Search  & Accuracy (\%)  & Number of Trees   & Projected Accuracy (\%)  & Avg. Per Search over all hashes  & Avg. Per Search   & \multicolumn{1}{r|}{Accuracy (\%) }\tabularnewline
\hline 
\multicolumn{6}{|l||}{LSH with RP-tree} & \multicolumn{2}{r|}{Angle rp-tree}\tabularnewline
\hline
\hline 
\multicolumn{8}{|>{\raggedright}p{1.5in}|}{mnist}\tabularnewline
\hline 
\multicolumn{1}{|>{\raggedright}p{0.5in}|}{12 } & 205.8  & 21  & 13  & 95.4  & 2675.4  & 10272.0  & \multicolumn{1}{r|}{94.9}\tabularnewline
\hline 
\multicolumn{1}{|>{\raggedright}p{0.5in}|}{10 } & 269.1  & 21.2  & 13  & 95.5  & 3498.3  & ''  & \multicolumn{1}{r|}{''}\tabularnewline
\hline 
\multicolumn{1}{|>{\raggedright}p{0.5in}|}{8 } & 613.8  & 27.4  & 10  & 96  & 6138  & ''  & \multicolumn{1}{r|}{''}\tabularnewline
\hline 
\multicolumn{1}{|>{\raggedright}p{0.5in}|}{4 } & 5855.3  & 51.7  & 4  & 94.6  & 23,421.2  & ''  & \multicolumn{1}{r|}{''}\tabularnewline
\hline 
\multicolumn{1}{|>{\raggedright}p{0.5in}|}{3 } & 9144.8  & 61.1  & 3  & 94.1  & 27,434.4  & ''  & \multicolumn{1}{r|}{''}\tabularnewline
\hline 
\multicolumn{8}{|l|}{}\tabularnewline
\multicolumn{8}{|>{\raggedright}p{1.5in}|}{bio\_train}\tabularnewline
\hline 
\multicolumn{1}{|>{\raggedright}p{0.5in}|}{14 } & 195.9  & 28.3  & 9  & 95  & 1763.1  & 3561.8  & \multicolumn{1}{r|}{93.9}\tabularnewline
\hline 
\multicolumn{8}{|l|}{}\tabularnewline
\multicolumn{8}{|l|}{bio\_test}\tabularnewline
\hline 
\multicolumn{1}{|>{\raggedright}p{0.5in}|}{14 } & 183  & 29.6  & 10  & 97.1  & 1830  & 3635.8  & \multicolumn{1}{r|}{97}\tabularnewline
\hline 
\multicolumn{8}{|l|}{}\tabularnewline
\multicolumn{8}{|l|}{Extended Yale Face B Cropped}\tabularnewline
\hline 
\multicolumn{1}{|>{\raggedright}p{0.5in}|}{9 } & 169.2  & 57.4  & 5  & 98.6  & 846  & 620.5  & \multicolumn{1}{r|}{99}\tabularnewline
\hline 
\multicolumn{8}{|l|}{}\tabularnewline
\multicolumn{8}{|l|}{Reuters Bag of Words}\tabularnewline
\hline 
\multicolumn{1}{|>{\raggedright}p{0.5in}|}{12 } & 217.9  & 33.3  & 5  & 86.8  & 1089.5  & 2917.8  & \multicolumn{1}{r|}{84}\tabularnewline
\hline 
\multicolumn{8}{|l|}{}\tabularnewline
\multicolumn{8}{|l|}{15D Sphere}\tabularnewline
\hline 
\multicolumn{1}{|>{\raggedright}p{0.5in}|}{17} & 192.3 & 13.2 & 19 & 93.2 & 3653.7 & 11,507 & \multicolumn{1}{r|}{93.2}\tabularnewline
\hline 
\multicolumn{8}{|l|}{}\tabularnewline
\multicolumn{8}{|l|}{20D Sphere}\tabularnewline
\hline 
\multicolumn{1}{|>{\raggedright}p{0.5in}|}{17} & 204.3 & 10.6 & 26 & 94.6 & 5311.8 & 20,757 & \multicolumn{1}{r|}{94.2}\tabularnewline
\hline
\end{tabular}
\par\end{centering}

~\\
\caption{Number of distance function calls for near neighbor search (only checking
one cell in the tree) for the mnist database. 1-NSS. This table compares
the angle rp-tree with LSH, where the hashing function is a series
of random projections.}

\label{Flo:LSHresults} 
\end{table*}
\par\end{center}

~~{\bf procedure} $createAngleTree(list of points pointList,~treetype)$
\begin{algorithmic}[1]
\STATE tree\_node node \COMMENT{Create node}
\IF{$size(pointList) \le minSize$} 
\STATE $node.data \gets pointsList$
\COMMENT{This will be a leaf node}
\ELSE
\STATE $node.splitter \gets genSplitter(treetype, pointList)$ \COMMENT{Generate splitter based on the treetype. This 
splitter will have a direction (its normal vector) and a threshold}
\STATE $node.angle \gets getAngle(node.splitter, pointList)$ \COMMENT {Our only modification to regular tree creation}
\STATE $node.negChild \gets createAngleTree(p \in pointList:~p \cdot node.normal \le threshold,~treetype )$ 
\STATE $node.negChild \gets angleTree(p \in pointList:~p \cdot node.normal > threshold,~treetype )$ 
\ENDIF
\end{algorithmic}
\vspace{3cm}
~~{\bf procedure} $getAngle(splitter,~pointList)$
\begin{algorithmic}[1]
\STATE $angles \gets [~]$
\STATE $center \gets $median (or mean) of pointList for each axis
\STATE $ret \gets 0$

\FOR[k is some not very large constant, say 2000]{$i = 1$ to k} 
\STATE $p \gets$ random point from pointList
\STATE $v \gets (p-center)$ \COMMENT{v is random vector within region containing pointList} 
\STATE $angles.append(\arccos\left(\frac{abs(<v,\, splitter.normal>)}{|v||splitter.normal|}\right)$ \COMMENT{angle between v and splitter.normal}
\ENDFOR
\STATE angles.sort()
\RETURN $angles[k(1-IOut)]$ \COMMENT{IOut is the Ignore Outlier parameter}

\end{algorithmic}
\vspace{1cm}
~~{\bf Pruning procedure during NNS}
\begin{algorithmic}[1]
\STATE $q \gets query~point$
\STATE $closestSoFar \gets dist(q,~closest~neighbor~found~so~far)$
\STATE We now modify our pruning criterion during NNS from:
\IF {$dist(q,~node.splitter) \ge closestSoFar$}
\STATE prune node's other child (where $q$ did not come from) 
\ENDIF
\STATE to:
\IF {$dist(q,~node.splitter)\cos\theta/\sin(node.angle) \ge closestSoFar$}
\STATE prune node's other child (where $q$ did not come from) 
\COMMENT {\ensuremath{\theta} is a small, constant Error Angle determined by k and d (see Section \ref{sec:estimDihed})}.  
\ENDIF
\end{algorithmic}

\section*{---------------------}


\begin{thebibliography}{40}
\bibitem{tanen}A global geometric framework for nonlinear dimensionality
reduction. \newblock J. B. Tenenbaum, V. De Silva and J. C. Langford
(2000). \newblock Science 290 (5500), 2319-2323.

\bibitem{Dasgupta}Freund, Yoav; Dasgupta, Sanjoy \newblock Random
projection trees and low dimensional manifolds.\newblock In Proceedings
of the 40th annual ACM symposium on Theory of computing, Pages 537-546,
2008.

\bibitem{Freund}Freund, Yoav; Dasgupta, Sanjoy; Kabra, Mayank; Verma,
Nakul \newblock Learning the structure of manifolds using random
projections\newblock NIPS, 2007

\bibitem{Investigation}Ting Liu, Andrew W. Moore, Alexander Gray
and Ke Yang\newblock An Investigation of Practical Approximate Nearest
Neighbor Algorithms \newblock NIPS, 2004

\bibitem{rptreeLect}Freund, Yoav; Dasgupta, Sanjoy \newblock Random
projection trees and low dimensional manifolds \newblock Lecture,
University of California, San Diego, http://www.stanford.edu/group/mmds/slides2008/dasgupta.pdf
accessed on 15/10/2009, 2008

\bibitem{Maneewongvatana1999}Songrit Maneewongvatana and David M.
Mount \newblock It's okay to be skinny, if your friends are fat \newblock
Center for Geometric Computing 4th Annual Workshop on Computational
Geometry, December 1999

\bibitem{Costa2000}Jose A. Costa and Alfred O. Hero III \newblock
Manifold Learning Using Euclidean K-Nearest Neighbor Graphs \newblock
Acoustics, Speech, and Signal Processing. Proceedings. (ICASSP '04).
IEEE International Conference, May 2004

\bibitem{Joly2008}Alexis Joly and Olivier Buisson\newblock A Posteriori
Multi-Probe Locality Sensitive Hashing{} \newblock International
Multimedia Conference, Proceeding of the 16th ACM international conference
on Multimedia, 2008

\bibitem{Andoni2008}Alexandr Andoni; Piotr Indyk \newblock Near-Optimal
Hashing Algorithms for Approximate Nearest Neighbor in High Dimensions
\newblock Communications of the ACM, SPECIAL ISSUE: Breakthrough
research: a preview of things to come, 2008

\bibitem{Andoni2009}Alexandr Andoni \newblock Nearest Neighbor Search:
the Old, the New, and the Impossible \newblock PhD thesis, Massachusetts
Institute of Technology 2009, September 2009

\bibitem{Beygel2006} Alina Beygelzimer; Sham Kakade; John Langford
\newblock Cover Trees for Nearest Neighbor \newblock ACM International
Conference Proceeding Series; Vol. 148, Proceedings of the 23rd international
conference on Machine learning, 2006

\bibitem{Comparison2005}Comparison of the Cover Tree and sb(S) Datastructures
Additional Experiments \newblock http://hunch.net/\textasciitilde{}jl/projects/cover\_tree/paper/
addendum/comparison.ps, January 2005

\bibitem{IndykLecture}Piotr Indyk \newblock Near-Optimal Hashing
Algorithms for Approximate Near(est) Neighbor Problem \newblock Lecture,
Massachusetts Institute of Technology, http://www.mit.edu/\textasciitilde{}andoni/papers/cSquared.pdf
accessed on{} 12/10/2009

\bibitem{IndykThesis}Piotr Indyk \newblock HIGH-DIMENSIONAL COMPUTATIONAL
GEOMETRY \newblock PhD thesis, Stanford University, September 2000

\bibitem{pami1995}Peter J. Verveer, Robert P.W. Duin \newblock \textquotedbl{}An
Evaluation of Intrinsic Dimensionality Estimators\textquotedbl{}
\newblock IEEE Transactions on Pattern Analysis and Machine Intelligence,
vol. 17, no. 1, pp. 81-86, Jan. 1995, doi:10.1109/34.368147

\bibitem{Lee77}D. T. Lee1 and C. K. Wong (1977) \newblock Worst-Case
Analysis for Region and Partial Region Searches in Multidimensional
Binary Search Trees and Balanced Quad Trees \newblock Acta Informatica,
Volume 9, Pages 23-29, 1977

\bibitem{Muja09}Marius Muja, David G. Lowe \newblock Fast Approximate
Nearest Neighbors with Automatic Algorithm Configuration \newblock
VISSAPP (1) 2009: 331-340

\bibitem{Shi06}Javen Qinfeng Shi \newblock Introduction to Cover
Tree \newblock Lecture, SML of NICTA RSISE of ANU October, http://users.rsise.anu.edu.au/\textasciitilde{}qshi/talk/introduction\%20to\%20covertree060815.pdf{} accessed on 10/11/2009,{}
2006

\bibitem{Yianilos2000}Peter N. Yianilos \newblock Locally lifting
the curse of dimensionality for nearest neighbor search \newblock
Symposium on Discrete Algorithms, Proceedings of the eleventh annual
ACM-SIAM symposium on Discrete algorithms,{} 2000

\bibitem{Lifshits07}Yury Lifshits \newblock Nearest Neighbors in
Doubling Metrics Algorithmic Problems Around the Web \#7 \newblock
Lecture, CalTech, CS101.2, http://yury.name/algoweb/07algoweb.pdf
accessed on 12/10/2009, Fall'07

\bibitem{Jordan09}Donghui Yan; Ling Huang; Michael I. Jordan \newblock
Fast Approximate Spectral Clustering \newblock International Conference
on Knowledge Discovery and Data Mining, Proceedings of the 15th ACM
SIGKDD international conference on Knowledge discovery and data mining,
2009

\bibitem{Weber}Roger Weber; Hans-Jorg Schek; Stephen Blott \newblock
A Quantitative Analysis and Performance Study for Similarity-Search
Methods in High-Dimensional Spaces \newblock Very Large Data Bases,
Proceedings of the 24rd International Conference on Very Large Data
Bases, 1998

\bibitem{Berinde08}R. Berinde and P. Indyk \newblock Sparse recovery
using sparse random matrices \newblock MIT-CSAIL Technical Report,
2008

\bibitem{JLLemma}S. Dasgupta and A. Gupta \newblock An elementary
proof of the Johnson\textendash{}Lindenstrauss lemma \newblock Technical
report 99\textendash{}006, U. C. Berkeley, March 1999

\bibitem{Hegde07}Chinmay Hegde; Richard G. Baraniuk; Michael B. Wakin
\newblock Random Projections for Manifold Learning \newblock NIPS,
December 2007

\bibitem{Verma07}Yoav Freund, Sanjoy Dasgupta, Mayank Kabra, Nakul
Verma \newblock Learning the structure of manifolds using random
projections \newblock NIPS,{} 2007

\bibitem{quantization08}Dasgupta, S. Freund, Y \newblock Random
projection trees for vector quantization \newblock Communication,
Control, and Computing, 46th Annual Allerton Conference, September
2008

\bibitem{Indyk04} Piotr Indyk \newblock Nearest Neighbors In High-Dimensional
Spaces \newblock Handbook of Discrete and Computational Geometry
(2nd edition) J. E. Goodman and J. O'Rourke, editors. CRC Press LLC,
2004

\bibitem{fisher}Ronald Fisher \newblock{} http://en.wikipedia.org/wiki/Chi-square\_distribution\#Asymptotic\_properties{}
accessed on 31/10/2009

\bibitem{Beis97}Beis, J. and Lowe, D. G. \newblock Shape indexing
using approximate nearest-neighbour search in high-dimensional spaces
\newblock Conference on Computer Vision and Pattern Recognition,
Puerto Rico, pp. 1000-1006, 1997

\bibitem{Karger2002}David R. Karger, Matthias Ruhl \newblock Finding
nearest neighbors in growth-restricted metrics \newblock Annual ACM
Symposium on Theory of Computing archive Proceedings of the thiry-fourth
ACM symposium on Theory of computing, 2002

\bibitem{mnist}The MNIST set of handwritten digits \newblock http://yann.lecun.com/exdb/mnist/ 

\bibitem{kdd}The 2004 KDD-cup data set \newblock http://kodiak.cs.cornell.edu/kddcup

\bibitem{yale1}Athinodoros Georghiades, Peter Belhumeur, and David
Kriegma \newblock From Few to Many: Illumination Cone Models for
Face Recognition under Variable Lighting and Pose \newblock PAMI,
2001

\bibitem{yale2}K.C. Lee and J. Ho and D. Kriegman \newblock Acquiring
Linear Subspaces for Face Recognition under Variable Lighting \newblock
IEEE Trans. Pattern Anal. Mach. Intelligence, Volume 27, Number 5,
Pages 684-698, 2005

\bibitem{multi-probe}Qin Lv, William Josephson, Zhe Wang, Moses Charikar,
Kai Li \newblock Multi-probe LSH: efficient indexing for high-dimensional
similarity search \newblock Very Large Data Bases; Proceedings of
the 33rd international conference on Very large data bases, 2007

\bibitem{friedman}Jerome H. Friedman, Jon Louis Bentley, Raphael
Ari Finkel \newblock An Algorithm for Finding Best Matches in Logarithmic
Expected Time \newblock ACM Transactions on Mathematical Software
(TOMS) Volume 3, Issue 3, September 1977

\bibitem{wikiIndyk}Jacob E. Goodman, Joseph O'Rourke and Piotr Indyk
(Ed.) (2004).\newblock \textquotedbl{}Chapter 39 : Nearest neighbors
in high-dimensional spaces\textquotedbl{}. \newblock Handbook of
Discrete and Computational Geometry (2nd ed.). CRC Press.

\bibitem{reutersBOW}Hettich, S. and Bay, S. D. (1999). \newblock
The UCI KDD Archive {[}http://kdd.ics.uci.edu{]}.\newblock Irvine,
CA: University of California, Department of Information and Computer
Science.

\bibitem{sift}R. M. Gray and D. L. Neuhoff \newblock Quantization
\newblock IEEE Transactions on Information Theory, vol. 44, pp. 2325\textendash{}2384,
Oct. 1998.

\bibitem{flann} Muja M, Lowe D. \newblock Fast approximate nearest neighbors with automatic algorithm configuration. \newblock Preprint. 2008. Available at: http://people.cs.ubc.ca/~lowe/papers/09muja.pdf. 

\bibitem{gnat} Brin S. \newblock Near neighbor search in large metric spaces. \newblock Proceedings of the International Conference on Very Large Data Bases.  \newblock INSTITUTE OF ELECTRICAL \& ELECTRONICS ENGINEERS (IEEE); 1995:574-584. Available at: http://scholar.google.com/scholar?hl=en\&btnG=Search\&q=intitle:Near+Neighbor+Search+in+Large+Metric+Spaces\#0.

\bibitem{dihed} Gellert, W.; Gottwald, S.; Hellwich, M.; Kästner, H.; and Künstner, H. (Eds.). \newblock VNR Concise Encyclopedia of Mathematics, 2nd ed. \newblock  New York: Van Nostrand Reinhold, 1989.

\bibitem{volCone} Su, Francis E., et al. \newblock "Volume of a Cone in N Dimensions." \newblock Mudd Math Fun Facts. <http://www.math.hmc.edu/funfacts>.

\bibitem{sphereCap} Harris, J. W. and Stocker, H. \newblock Spherical Segment (Spherical Cap) 
\newblock §4.8.4 in Handbook of Mathematics and Computational Science. New York: Springer-Verlag, p. 107, 1998.


\bibitem{gaussFunc} Weisstein, Eric W. "Hypergeometric Function." From MathWorld--A Wolfram Web Resource. http://mathworld.wolfram.com/HypergeometricFunction.html


\bibitem{chiDistrib} Weisstein, Eric W. "Chi Distribution." From MathWorld--A Wolfram Web Resource. http://mathworld.wolfram.com/ChiDistribution.html

\end{thebibliography}
\end{document}